# Stabilized Modulated Photonic Signal Transfer Over 186 km of Aerial Fiber

David R. Gozzard,[1,2,*] Sascha W. Schediwy,[2,1] Bruce Wallace,[3] Romeo Gamatham,[4] and Keith Grainge[5]

[1]*School of Physics and Astrophysics, University of Western Australia, Perth, WA 6009, Australia*
[2]*International Centre for Radio Astronomy Research, University of Western Australia, Perth, WA 6009, Australia*
[3]*SKA South Africa, 17 Baker Street, Rosebank, Johannesburg, South Africa*
[4]*SKA South Africa, 3rd Floor, The Park, Park Road, Pinelands 7405, South Africa*
[5]*Jodrell Bank Centre for Astrophysics, Alan Turing Building, School of Physics & Astronomy, The University of Manchester, Oxford Road, Manchester M13 9PL, UK*

*\*david.gozzard@research.uwa.edu.au*

**Abstract:** Aerial suspended optical-fiber links are being considered as economical alternatives to buried links for long-distance transfer of coherent time and frequency signals. We present stability measurements of an actively stabilized 20 MHz photonic signal over aerial fiber links up to 186.2 km in length. Absolute frequency stabilities of $2.7 \times 10^{-3}$ Hz at 1 s of integration, and $2.5 \times 10^{-5}$ Hz at $8 \times 10^3$ s of integration are achieved over this longest link. This stability is compared to that achieved over buried links for both radio and microwave frequencies. The results show that aerial fiber links are a suitable alternative to buried links for a wide range of frequency transfer applications.

## 1. Introduction

Stabilized frequency transfer over long-distance fiber-optic networks has applications in science and industry ranging from radio astronomy, earth sciences, and tests of fundamental physics, to network synchronization and monitoring [1]. As stabilized frequency transfer technologies have matured in recent years, they have rapidly moved beyond the laboratory in to real-world applications [2-5].

Studies of the frequency transfer stability over in-the-field optical fiber installations have focused on underground [6-8] and submarine [9] links, as well as telecommunications links incorporating dense wavelength division multiplexing (DWDM) infrastructure [6-8]. Frequency transfer stability over aerial suspended fiber cables and optical ground wires (OPGWs) has received less attention because their greater exposure to environmental perturbations means that frequency transfer stability over these links is severely degraded compared with buried links. Overhead links have been shown to exhibit orders-of-magnitude more noise than underground links of similar length [10, 11] which represents a major challenge for frequency stabilization systems. As a result, most research on overhead links has focused on characterizing polarization mode dispersion, state of polarization fluctuations, and transmission delay variations and their impact on telecommunications and network monitoring [12-16]. However, the greatly reduced cost of reticulating overhead fiber links relative to underground links, as well as the prevalence of existing overhead fiber infrastructure (usually in the form of OPGWs), means that there are situations in which it is necessary or desirable to exploit aerial fiber links for stable frequency transmission. Specifically, the tests reported in this paper were carried out in order to determine the effectiveness of disseminating stabilized frequency references over aerial fiber links that will form the backbone of the Synchronization and Timing network for the component of the Square Kilometre Array radio telescope to be constructed in the Karoo region of South Africa (SKA-mid) [17, 18].

In this paper, we report the successful stabilization of a 20 MHz modulated photonic signal transmitted over aerial suspended fiber links of 32.6 km, 153.6 km, and 186.2 km in length, and which were subject to adverse weather conditions including high wind speeds and large thermal gradients. The transferred frequency stabilities of these links are compared to the performance of transferred frequency stabilities over buried links of similar length. Results in terms of absolute frequency stability are used to extend the discussion of the practicality of aerial links across multiple frequency bands.

## 2. Experimental design

These tests were conducted at SKA South Africa's Karoo support base at Klerefontein. The links used in these tests comprised two cores of a 16.3 km aerial suspended fiber cable between Klerefontein and the Carnarvon point-of-presence (POP) site, and two cores of a 76.8 km aerial suspended fiber cable between Klerefontein and the SKA site. Figure 1 is a map depicting the relevant locations and fiber routes.

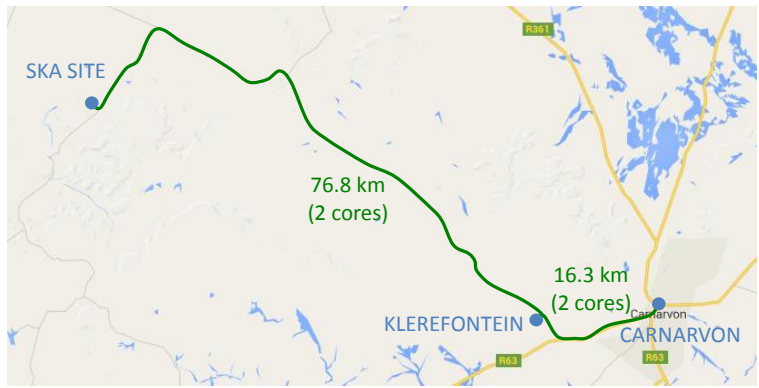

Fig. 1. Map showing locations and fiber link routes relevant to the described tests. (Modified from Google Maps image.)

A fiber patch lead was installed at the Carnarvon POP site to produce a loop-back link of 32.6 km. To compensate for the large optical power loss on the 76.8 km link (31.7 dB loss), a patch incorporating an IDIL Fibres Optiques bi-directional optical amplifier (with a gain of +18.7 dB and a noise figure of 7.6 dB) was installed at the SKA site to produce a loop-back length of 153.6 km. The fiber in the aerial cables was of standard SMF E9, according to ITU-T G.652.D, with a chromatic dispersion coefficient of 18 ps/nm·km at a wavelength of 1550 nm. The aerial fiber cables were suspended from utility poles for the entirety of their runs and had a typical span distance of 70 m.

Figure 2 shows a schematic representation of the stabilized signal transfer system deployed at the Karoo support base at Klerefontein. This stabilized modulated photonic signal transmission system is based on that presented in [19] which achieves phase-noise suppression of the modulated photonic signal through optical phase-sensing and optical phase-actuation using acousto-optic modulators (AOMs). The transmitter and receiver of the stabilized transfer system were co-located in the same room at the support base to enable out-of-loop measurements of the transfer stability to be made.

In the transmitter section of the system, an NKT Photonics Koheras BASIK X15 commercial diode laser (spectral linewidth < 100 Hz) was used to provide a highly coherent optical frequency at 193 THz (1552 nm). The optical signal was split and part of the signal was reflected off of a Faraday mirror (FM) and onto the transmitter-side photodetector (PD) to provide the optical reference signal for an imbalanced Michaelson interferometer. The other part of the optical signal from the laser entered a Mach-Zehnder interferometer where a pair of AOMs shifted the optical frequency by 50 MHz and 70 MHz, producing a single-sideband modulated 20 MHz signal at the output of the Mach-Zehnder. This modulated photonic signal was injected into the aerial fiber link.

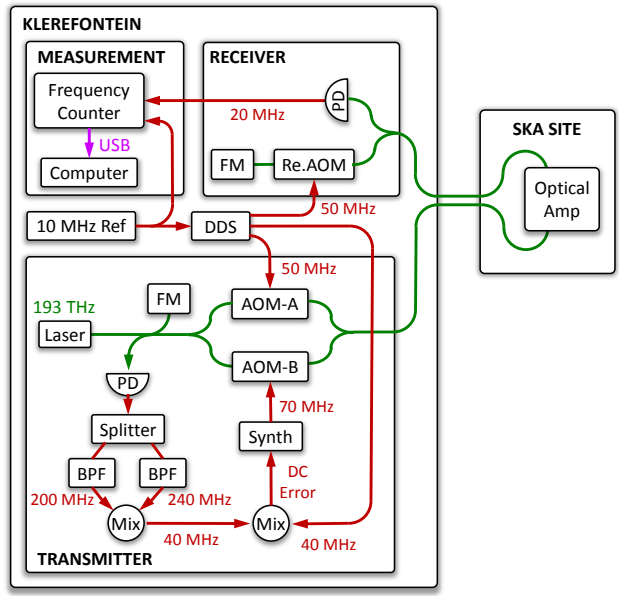

Fig. 2. Schematic of the stabilized modulated photonic signal transfer system and 2×76.8 km fiber link with optical amplifier. PD, photodetector; AOM, acousto-optic modulator; FM, Faraday mirror; BPF, band-pass filter; Mix, frequency mixer; DDS, direct digital synthesizer.

The transmitted signal made the round trip through the link and entered the receiver section of the system. The optical signal was split with part being directed to the measurement PD. The 20 MHz signal from the measurement PD was counted directly by an Agilent 53132A high-precision frequency counter.

The rest of the received optical signal passed through the receiver AOM which introduced a 2×50 MHz frequency shift so that the optical signals returning to the transmitter could be distinguished from reflections on the fiber link. The optical signals reflected off of a FM and returned through the link to the transmitter where they again passed through the Mach-Zehnder interferometer and entered the transmitter-side PD where they beat against the optical reference to produce electronic signals.

The signals of interest were 200 MHz (the signal that double-passed the 50 MHz AOM in the Mach-Zehnder) and 240 MHz (the signal that double-passed the 70 MHz AOM). These electronic beat signals encoded information about the frequency fluctuations resulting from environmental perturbations on the link. The signals were split and filtered separately to remove the unwanted frequencies. The 200 MHz and 240 MHz signals were then mixed, producing a 40 MHz signal that was then mixed with a 40 MHz local oscillator to produce a DC error signal that was fed to the synthesizer supplying the 70 MHz AOM. When the control loop was closed, the DC error signal adjusted the frequency of the 70 MHz AOM to suppress frequency fluctuations on the link. (A full mathematical description of the operation of this type of stabilized modulated photonic signal transfer system is given in [19].) Measurements of stabilized transmissions were performed over links of 32.6 km (Klerefontein—Carnarvon POP—Klerefontein, 8.6 dB loss), 153.6 km (Klerefontein—SKA site—Klerefontein, 31.7 dB loss, 18.7 dB gain from optical amplifier), and 186.2 km (Klerefontein—SKA site—Klerefontein—Carnarvon POP—Klerefontein, 40.3 dB loss, 37.4 dB gain from two optical amplifiers). Optical amplifiers were used as necessary to compensate for the attenuation of the optical signal on the links. The 153.6 km link used the one amplifier installed at the SKA-site, while the 186.2 km link incorporated a second amplifier at Klerefontein which served to boost the signal arriving from the SKA site before it was directed to the Carnarvon POP. Measurements of stabilized transmission over the two longest links were made for periods of

48 hours, while the transmission over the 32.6 km link was measured for one hour. The measurements of the 32.6 km link were run for a shorter period of time than for the longer links because the primary goal of the tests was to determine the ability of the frequency transfer system to stabilized the longer links. The 32.6 km link was measured only for comparison.

To assess the ability of the stabilization system to cope with adverse weather conditions, weather data, including wind speed and air temperature, coinciding with the period of the tests were collected from a weather station operated by the C-Band All Sky Survey (C-BASS) at the Klerefontein site.

## 3. Results

The frequency counter data were processed to produce a triangle-weighted estimate of the absolute frequency stability of the transmitted photonic signals. Results are presented in terms of absolute frequency stability because this more readily allows for the comparison of different transmission frequencies that is covered in the discussion section. Unlike fractional frequency stability, absolute frequency stability is independent of the value of the transmitted frequency and, as will be demonstrated in the discussion section, allows a prediction to be made of the performance of different stabilized frequency transmission systems over these aerial links. Fractional frequency stability values are readily obtained by dividing the absolute frequency stability by the transmission frequency (in this case, 20 MHz). The absolute frequency stability curves are shown in Figure 3 (a). Figure 3 (b) shows the measured absolute frequency stability of the 20 MHz transmission over "ground" links that were within 7% of the length of the aerial links. These ground-comparison tests were performed in the laboratory in Perth, Western Australia using the same frequency transfer system on links that comprised a mix of metropolitan networks and spooled fiber up to 144 km in length. To improve the comparison, the noise levels measured on the ground links were scaled using the procedure developed by Williams and colleagues [20] to estimate the noise level on ground links of exactly the same length as the aerial links. This was achieved by multiplying the measured absolute frequency stability of the ground link by $(L2/L1)^{3/2}$, where L1 is the length of the measured ground link and L2 is the length of the link for which the prediction is being made.

The black traces in Figure 3 (a) and (b) (black triangles) are 'zero-length' stability measurements made by connecting the transmitter and receiver with a fiber patch lead with attenuation set equal to that of the 32.6 km aerial link (8.6 dB loss). As such, these traces represent the noise floor of the transmission system. The discrepancy between the noise floor measurements for the aerial and ground links is likely due to the lack of environmental shielding for the stabilization system at the Kelerefontein site. The server room at Klerefontein in which the stabilization system was installed is a re-purposed farm house with very poor insulation and draft-exclusion. This subjected the stabilization system to a significant fraction of the large and rapid temperature fluctuations shown in Figure 4, impacting the profile of the measured noise floor. In contrast, the ground link tests were performed in a temperature-controlled metrology laboratory.

The 32.6 km aerial transmission (Fig. 3 (a), green diamonds) achieved an absolute frequency stability of $4.2 \times 10^{-4}$ Hz at one second of integration time. Using the scaling equation described previously [20], the measured absolute frequency stability of the 31 km ground link (31 km of metropolitan fiber network, Fig. 3 (b), green, filled diamonds, solid line) was extrapolated to predict the stability of an equivalent 32.6 km ground link (Fig. 3 (b), green, open diamonds, dashed line). The predicted absolute frequency stability for this link at 1 s of integration time is $3.1 \times 10^{-4}$ Hz, slightly better than the stability of the aerial link. The stability of this predicted ground link drops to $3.1 \times 10^{-6}$ Hz at $8 \times 10^3$ s of integration time.

The aerial 153.6 km transmission (Fig. 3 (a), orange squares) achieved an absolute stability of $1.5 \times 10^{-3}$ Hz at one second of integration time, dropping to $8.9 \times 10^{-6}$ Hz at $8 \times 10^3$ s of integration time. The longest available ground link for comparison was 144 km in length (31 km metropolitan network plus 123 km of spooled fiber). The measured frequency stability of this link (Fig. 3 (b), orange, filled squares, solid line) was extrapolated to predict the stability of a ground link of 153.6 km in length (Fig. 3 (b), orange, open squares, dashed line). The resulting predicted absolute stability was $1.1\times10^{-3}$ Hz at one second of integration time dropping to $6.2\times10^{-6}$ Hz at $8\times10^3$ s of integration time.

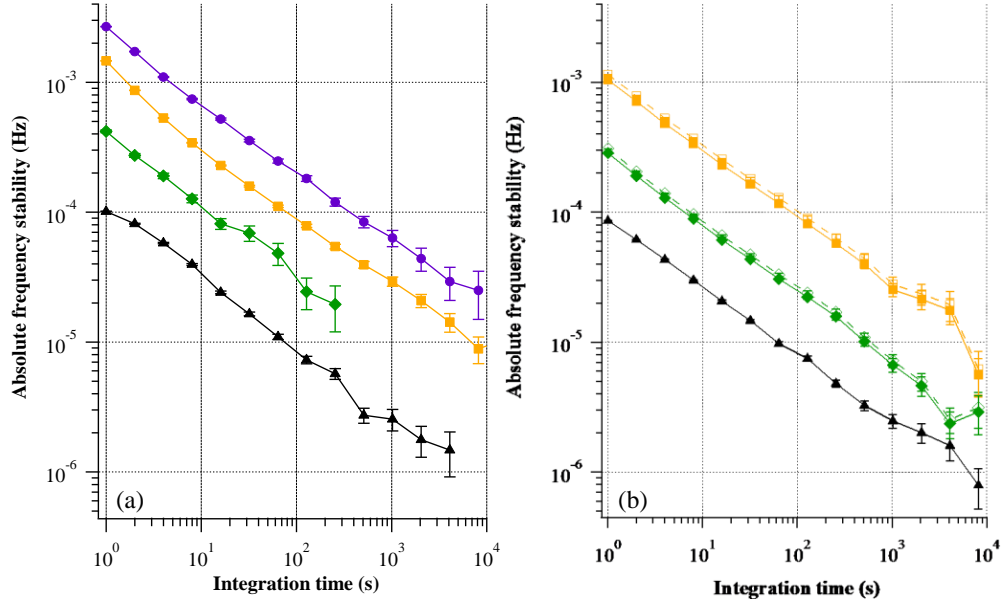

Fig. 3. Absolute frequency stabilities for (a) aerial and (b) ground links derived from the Agilent 53132A frequency counter. Trace styles indicate aerial and ground links of similar lengths. Traces with open markers and dashed lines indicate extrapolated data. Green, filled diamonds, solid line —(a) 32.6 km, (b) 31 km; green, open diamonds, dashed line — 31 km link extrapolated to 32.6 km; orange, filled squares, solid line — (a)153.6 km, (b) 144 km; orange, open squares, dashed line — 144 km link extrapolated to 153.6 km; purple circles — 186.2 km; black triangles — 'zero-length' noise floor.

The 186.2 km transmission (Fig. 3 (a), purple circles) achieved an absolute stability of $2.7\times10^{-3}$ Hz at one second of integration time, and $2.5\times10^{-5}$ Hz at $8\times10^3$ s of integration time. Because this 186.2 km aerial link was nearly 30% longer than the most similar length ground link (144 km), no aerial-ground comparison was made for these measurements.

The results in Figure 3 show that the transfer stabilities over aerial and ground links of comparable lengths are very similar, with the aerial links displaying frequency stability values that were only around 35% higher than the ground links despite the large stress loadings imposed by climatic effects.

The frequency transfer over the aerial links proved to be very robust, with the system demonstrating its ability to stabilize an aerial link subject to a wide variety of weather conditions. The measurements of the 32.6 km link were performed during stormy weather in which the speed of wind gusts exceeded 60 km/h. The period corresponding to the measurements of the 153.6 km and 186.2 km links had weather conditions that varied from thick overcast, with wind gust speeds around 50 km/h, to fine conditions with wind gusts not exceeding 28 km/h. The SKA's operational requirements rule that non-weather protected equipment and fiber must operate at wind speeds up to 40 km/h, and over a temperature range of −5°C to 50°C with a maximum temperature gradient of ±3°C over 10 minutes. Figure 4 shows wind gust speed and air temperature data from the C-BASS weather station. The time periods corresponding with frequency transfer measurements are shaded and labeled.

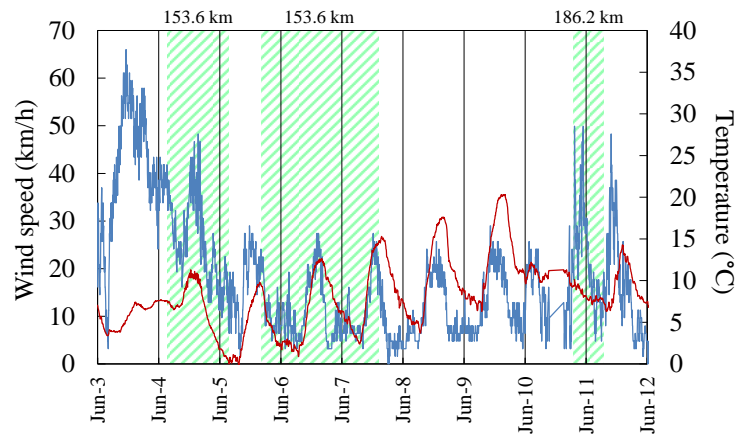

Fig. 4. Wind gust speed (blue line) and air temperature (red line) as recorded at the Klerefontein site for the period coinciding with the tests performed on 153.6 km and 186.2 km links. The shaded regions indicate the times when frequency transfer measurements were being performed.

Figure 4 shows that measurements of stabilized transfer over the 153.6 km and 186.2 km links were made during periods of high winds, demonstrating that the stabilized transfer system could maintain a robust lock despite the large stresses induced on the aerial cable spans and at wind speeds in excess of the SKA operational requirements. Due to the thick overcast, the thermal gradients on the cables during these times were relatively small compared to the days of fine weather and little overcast. Fine days produced greater thermal gradients due to the larger air-temperature range and the effects of solar heating of the aerial cables. In particular, frequency fluctuations indicative of thermal expansion of the aerial cables were most apparent in the data corresponding to periods during the morning after sunrise, when the cables went from being subjected to overnight frost to direct sunlight. The stabilized signal transfer system was able to operate continuously throughout these weather conditions. No cycle slips were recorded for the 24 hours of the 153.6 km link measurements or the 16 hours of the 186.2 km link measurements coinciding with the periods of higher wind speeds (see Fig. 4). Also, no cycle clips were recorded for 48 hours of the 153.6 km link measurement coinciding with the period of larger thermal gradients (see Fig. 4). However, these weather conditions were not extreme enough to subject the fiber link and stabilization system to the temperature range and gradient limits required by the SKA. During the measurements, the air temperature ranged from 0°C to 15°C with a maximum thermal gradient of 1.2°C in 10 minutes.

### 4. Discussion

The choice of transmission frequency (20 MHz) for these tests was determined mainly by the AOMs available at the time. Higher transmission frequencies would improve the fractional frequency stability performance of the system. Tests of this type of optically-sensed/actuated modulated photonic signal transfer system [19] at different transmission frequencies over other fiber links have shown that the absolute frequency stability of the system is largely independent of the transmitted frequency. This effect is demonstrated in Figure 5.

Figure 5 shows the absolute frequency stability of the 20 MHz transfer over 144 km of ground fiber extrapolated to the predicted performance of a 166 km link (orange, open squares, dashed line). This is compared to 160 MHz transfer over 166 km of metropolitan fiber link (red circles) using the modulated photonic signal transfer system from [19] on which the 20 MHz transfer system is based, and to 8,000 MHz transfer over the 166 km link (blue triangles) using a related design (the 8,000 MHz data is taken from Figure 2 in [5]). Despite the fact that the highest and lowest frequencies differ by a factor of 400, the absolute frequency transfer stabilities are within less than a factor of three of each other. The remaining differences in the absolute stability levels shown by the three traces can be attributed to the fact that the experimental setups were not identical. Although based on the same design, the 20 MHz and 160 MHz transmission systems used different models of photodetectors, filters, mixers, and other electronics, and transmitted over a differently configured link. The design of the 8,000 MHz stabilized transfer system, although related to the design transmitting the lower frequencies, utilized substantially different electronics to accommodate the higher frequencies involved. The independence of the absolute frequency stability from the transmitted frequency value means that the fractional frequency stability increases with increasing transmission frequency.

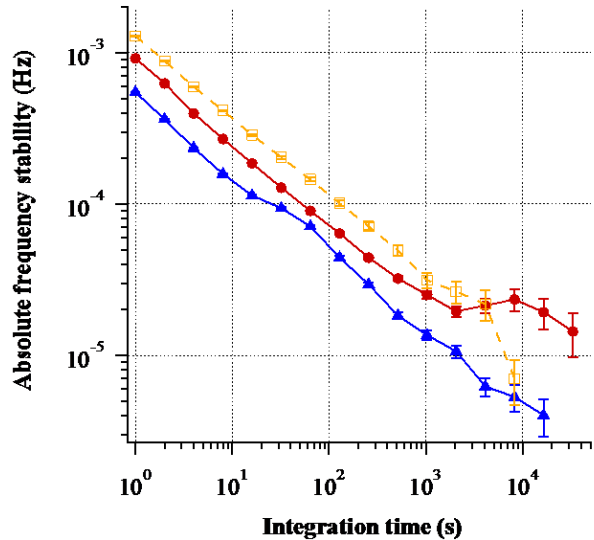

Fig. 5. Absolute frequency stability of transmissions over 166 km of buried fiber at 20 MHz (orange, open squares, dashed line), 160 MHz (red circles), and 8,000 MHz (blue triangles, data from [5] Fig. 2).

The stability of the transmission along the whole link can also be improved by the use of repeater stations at intervals along the link. This increases the servo bandwidth and enables greater servo gain and greater suppression of short-timescale noise processes. Assuming the stabilized frequency transfer systems in the repeater stations are configured optimally and an idealized link in which all link segments experience the same level of environmental noise, for a link of length L divided into N equal segments, the frequency stability of each segment will be a factor of $(1/N)^{3/2}$ times the frequency stability of the entire link as a single span (using the link noise scaling rule from [20]). The combined frequency stability of these N link segments in series is $\sqrt{N}$ times the frequency stability of a single segment. Thus, splitting a fiber link into N individually stabilized segments produces an overall improvement in the frequency stability at the remote site of 1/N (discounting the effects of adding devices such as optical amplifiers). For example, if the 153.6 km aerial fiber link considered in Figure 3 were split into two 76.8 km segments with separate stabilized frequency transfer systems, the absolute frequency stability of the 20 MHz transfer would be improved by a factor of 2, from $1.5\times10^{-3}$ Hz at one second of integration time to $7.5\times10^{-4}$ Hz at one second of integration time.

## 5. Conclusions

We have demonstrated the stabilized transfer of a 20 MHz photonic signal over aerial suspended fiber links up to 186.3 km in length. The frequency transfer proved to be very robust, demonstrating continuous stabilized transfer throughout inclement weather conditions including high winds and large thermal gradients. The overall stability levels measured on the aerial links were comparable to ground links of similar length using the same frequency transfer system. Increasing the modulation frequency offers an immediate way to significantly improve the fractional frequency stability levels attained by this type of stabilized photonic signal transfer system [19], and the addition of repeater stations at intervals along the link can also improve the performance of the whole link.

Based on the overall noise levels observed for all of the transmissions, is seems likely that other modulated photonic signal frequency transfer systems, such as [21], [22], and [23], will also achieve effective stabilization on aerial links. However, the infinite feedback range provided by AOM-actuated systems offers advantages over the group-delay actuator technology of [21] and [22] over long links subjected to large thermal gradients. These results indicate that aerial fiber is a suitable medium for stabilized frequency transfer and presents an attractive alternative when underground links are unavailable or prohibitively expensive. The independence of the absolute frequency stability demonstrated in Figure 5 allows the performance of similar modulated photonic signal frequency transfer systems operating at different frequencies on the overhead links to be predicted. Early results from another set of tests carried out on these South African links on behalf of the SKA [24] indicate that aerial links are also suitable for microwave frequency transfer.

## Funding

University of Manchester, University of Western Australia (UWA).

## Acknowledgement

We are very grateful to Roufurd Julie and Johan Burger for their efforts in supporting this field trip, and to Charles Copley for providing the C-BASS weather station data. This paper describes work being carried out for the SKA Signal and Data Transport (SaDT) consortium as part of the Square Kilometre Array (SKA) project. The SKA project is an international effort to build the world's largest radio telescope, led by the SKA Organisation with the support of 10 member countries.